\documentclass{emulateapj}

\usepackage{epsfig}

\newcommand{\etal}{\mbox{et al.}}
\newcommand{\ergcms}{erg cm$^{-2}$ s$^{-1}$}
\newcommand{\ergs}{erg s$^{-1}$}
\newcommand{\phcms}{ph cm$^{-2}$ s$^{-1}$}
\newcommand{\degree}{$^\circ$}

\newcommand{\chandra}{{\it Chandra}}

\newcommand{\bepposax}{{\it BeppoSAX}}
\newcommand{\xmm}{{\it XMM-Newton}}

\newcommand{\rxte}{{\it RXTE}}

\newcommand{\sgrastar}{\mbox{Sgr A$^*$}}
\newcommand{\newsource}{CXOGC~J174540.0--290031}

\makeatletter

\newcommand\newtablebreak{\cr\ptable@@split}
\makeatother

\shortauthors{Muno \etal}
\shorttitle{A Remarkable Galactic Center Transient}

\begin{document}
\title{A Remarkable Low-Mass X-ray Binary within 0.1 pc of the Galactic Center}
\author{M. P. Muno,\altaffilmark{1,2} 
J. R. Lu,\altaffilmark{1}
F. K. Baganoff,\altaffilmark{3} 
W. N. Brandt, \altaffilmark{4}
G. P. Garmire,\altaffilmark{4}
A. M. Ghez,\altaffilmark{1}
S. D. Hornstein,\altaffilmark{1} \\ 
\& M. R. Morris\altaffilmark{1}
}

\altaffiltext{1}{Department of Physics and Astronomy, University of California,
Los Angeles, CA 90095; mmuno@astro.ucla.edu}
\altaffiltext{2}{Hubble Fellow}
\altaffiltext{3}{Kavli Institute for Astrophysics and Space Research,
Massachusetts Institute of Technology, Cambridge, MA 02139}
\altaffiltext{4}{Department of Astronomy and Astrophysics, 
The Pennsylvania State University, University Park, PA 16802}

\begin{abstract}
Recent X-ray and radio observations have identified a transient low-mass 
X-ray binary (LMXB) located only 0.1 pc in projection from the 
Galactic center, \newsource\ \citep{mun05,bow05}. In this paper, we report 
the detailed analysis of X-ray and infrared observations of the transient 
and its surroundings. \chandra\ observations detect the source at a flux of 
$F_{\rm X} = 2\times10^{-12}$~\ergcms\ (2--8 keV). After accounting for 
absorption both in the interstellar medium and in material local to the 
source, the implied luminosity of the source is only 
$L_{\rm X} = 4\times10^{34}$~\ergs\ (2--8~keV; $D = 8$~kpc).  
However, the diffuse X-ray emission near the source also brightened by 
a factor of 2. The enhanced diffuse X-ray emission lies on top of a known
ridge of dust and ionized gas that is visible
infrared images. We interpret the X-ray emission as scattered flux from 
the outburst, and determine that the peak luminosity of \newsource\
was $L_{\rm X} \ga 2\times10^{36}$~\ergs. 
We suggest that the relatively
small observed flux results from the fact that the system is observed
nearly edge-on, so that the accretion disk intercepts most of the 
flux emitted along our line of sight.
We compare the inferred peak X-ray luminosity to that of the radio jet. 
The ratio of the X-ray to radio luminosities, 
$L_{\rm X}/L_{\rm R} \la 10^4$, is considerably smaller than in other 
known LMXBs ($\ga 10^5$). This is probably because the jets are 
radiating with unusually high efficiency at the point where they impact
the surrounding interstellar medium. This hypothesis is supported 
by a comparison with mid-infrared images of the surrounding dust. 
Finally, we find that the minimum power required to produce the 
jet, $L_{\rm jet} \sim 10^{37}$~\ergs, is comparable to the
inferred peak X-ray luminosity. This is the most direct 
evidence yet obtained that LMXBs accreting at low rates release about half
of their energy as jets. 
\end{abstract}


\section{Introduction}

Black holes and neutron stars accreting from binary companions
are often identified as transient X-ray, radio, 
optical, and infrared sources. Most of the identifications have occurred
first at X-ray wavelengths, thanks to a series space-based observatories
that were designed to monitor large portions of the sky 
\citep[e.g.][]{lev96,jag97}. However, the sensitivities of these instruments
are only $\simeq$$10^{-10}$~\ergcms, which at the Galactic center
distance corresponds to $\simeq$$10^{36}$~\ergs\ 
\citep[$D$=8~kpc; see][]{rei99}.

The \chandra\ X-ray Observatory and \xmm\ are several 
orders of magnitude more sensitive than previous wide-field 
X-ray instruments. Their observations of large concentrations of 
stars in our Galaxy, such a globular clusters and the Galactic center, have
revealed several faint X-ray transients with 
$L_{\rm X}$$\sim$$10^{34}$--$10^{35}$ \ergs
\citep[e.g.,][]{hei03,sak04,mun05}. 
Studies of these faint X-ray transients provide
the only means of understanding the physics of accretion at luminosities 
between the more well-examined regimes of outburst 
($L_{\rm X} > 10^{36}$ \ergs) and quiescence ($L_{\rm X} < 10^{33}$ \ergs).

In 2004 July, \chandra\ observations of the Galactic Center 
revealed a new transient X-ray source, \newsource, that was 
located only 0.1~pc in projection from the super-massive black 
hole \sgrastar. The close 
proximity of the source to \sgrastar\ has important implications for 
understanding stellar dynamics at the Galactic center, 
which have been discussed in \citet{mun05}.
In that paper, we also reported a candidate 7.9~h 
orbital modulation in the X-ray light curve, and an upper limit
on the magnitude of any infrared counterpart of 
$K$$<$16 in Keck obsevrations. These facts demonstrated that the 
source is a low-mass X-ray binary (LMXB). 
This source also was detected with \xmm\ in 2004 March and August, as is 
discussed in \citet{bel05} and D. Porquet \etal, (in prep.).
Finally, \newsource\ was detected as a radio transient with the Very
Large Array during a series of obervations between 2004 March and 2005 
January. A detailed analysis of the radio emission is presented in 
\citet{bow05}.

In this paper, we provide the details of our analysis of the \chandra\ and
Keck observations. We also report that the diffuse X-ray emission within 
3\arcsec\ of \newsource\ has brightened. We interpret this 
diffuse feature as a light echo from the transient, which allows us
to constrain the peak luminosity of the outburst. Finally, 
we compare the energetics of the X-ray and radio outburst, in order to 
understand how accretion proceeds in this remarkable example of a faint
X-ray transient.

\begin{deluxetable*}{lccccc}[thp]
\tablecolumns{6}
\tablewidth{0pc}
\tablecaption{Observations of the Inner 20 pc of the Galaxy\label{tab:obs}}
\tablehead{
\colhead{} & \colhead{} & \colhead{} & 
\multicolumn{2}{c}{Aim Point} & \colhead{} \\
\colhead{Start Time} & \colhead{Sequence} & \colhead{Exposure} & 
\colhead{RA} & \colhead{DEC} & \colhead{Roll} \\
\colhead{(UT)} & \colhead{} & \colhead{(s)} 
& \multicolumn{2}{c}{(degrees J2000)} & \colhead{(degrees)}
} 
\startdata
1999 Sep 21 02:43:00 & 0242  & 40,872 & 266.41382 & $-$29.0130 & 268 \\
2000 Oct 26 18:15:11 & 1561 & 35,705 & 266.41344 & $-$29.0128 & 265 \\
2001 Jul 14 01:51:10 & 1561 & 13,504 & 266.41344 & $-$29.0128 & 265 \\
2002 Feb 19 14:27:32 & 2951  & 12,370 & 266.41867 & $-$29.0033 & 91 \\
2002 Mar 23 12:25:04 & 2952  & 11,859 & 266.41897 & $-$29.0034 & 88 \\
2002 Apr 19 10:39:01 & 2953  & 11,632 & 266.41923 & $-$29.0034 & 85 \\
2002 May 07 09:25:07 & 2954  & 12,455 & 266.41938 & $-$29.0037 & 82 \\
2002 May 22 22:59:15 & 2943  & 34,651 & 266.41991 & $-$29.0041 & 76 \\
2002 May 24 11:50:13 & 3663  & 37,959 & 266.41993 & $-$29.0041 & 76 \\
2002 May 25 15:16:03 & 3392  & 166,690 & 266.41992 & $-$29.0041 & 76 \\
2002 May 28 05:34:44 & 3393  & 158,026 & 266.41992 & $-$29.0041 & 76 \\
2002 Jun 03 01:24:37 & 3665  & 89,928 & 266.41992 & $-$29.0041 & 76 \\
2003 Jun 19 18:28:55 & 3549 & 24,791 & 266.42092 & --29.0105 & 347 \\
2004 Jul 05 22:33:11 & 4683 & 49,524 & 266.41605 & --29.0124 & 286 \\
2004 Jul 06 22:29:57 & 4684 & 49,527 & 266.41597 & --29.0124 & 285 \\
2004 Aug 28 12:03:59 & 5630 &  5,106 & 266.41477 & --29.0121 & 271 \\
2005 Feb 27 06:26:04 & 6113 &  4,855 & 266.41870 & --29.0035 & 91 
\enddata
\end{deluxetable*}

\begin{figure*}[thb]
\centerline{\epsfig{file=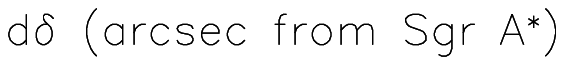,angle=90,width=0.8\linewidth}}
\caption{
Images of the 10\arcsec\ around the super-massive black hole \sgrastar, 
which illustrate the appearance of \newsource. 
{\it Left:} Image created from the average of 13 observations (650~ks
exposure) taken between 1999 September and 2003 June, demonstrating the 
quiescent state of the region. {\it Right:} Image created from 2 observations
(99~ks) taken on 2004 July 5--7, in which a new transient X-ray source is
evident 2\farcs5 south of \sgrastar\ (circle). The location of
twin lobes of the radio transient identified with the VLA are indicated by 
diamonds. Finally,
a portion of the diffuse emission brightened coincident with the transient
outburst (ellipse). Both images are displayed at the 0\farcs5 resolution of 
the detector. They were generated from the
raw counts, and then scaled to correct for the relative exposures and 
the spatially-varying effective area of the detector.
}
\label{fig:img}
\end{figure*}

\newpage

\section{X-ray Observations}

The \chandra\ X-ray Observatory has observed the inner 10\arcmin\ of 
the Galaxy with the Advanced CCD Imaging Spectrometer 
imaging array \citep[ACIS-I;][]{wei02} at least once a year between 
1999 and 2004 \citep[Table~\ref{tab:obs};][]{bag03,mun03a,mun05}. 
As mentioned in \citet{mun05}, a new transient source, \newsource,
 was identified 2\farcs9 south of 
\sgrastar\ during 99~ks of observations on 2004 July 5--7 
(Fig.~\ref{fig:img}), and during 5~ks of director's discretionary 
observations on 2004 August 28. We obtained another 5 ks 
observation of the field on 2005 February 27, which we report here for 
the first time. 

Each observation has been processed using the techniques described in 
\citet{mun03a}. In brief, for each observation we corrected the pulse 
heights of the events for
position-dependent charge-transfer inefficiency \citep{tow02b}, excluded 
events that did not pass the standard ASCA grade filters and \chandra\ X-ray
center (CXC) good-time 
filters, and removed intervals during which the background rate flared to
$\ge 3\sigma$ above the mean level. Finally, we applied a correction to 
the absolute astrometry of each pointing using three Tycho sources detected 
strongly in each \chandra\ observation \citep[][]{bag03}. We estimated
combined accuracy of our astrometric frame and of the positions of the 
individual X-ray sources by comparing the offsets between 36 foreground X-ray 
sources 
that were located within  5\arcmin\ of 
\sgrastar\ \citep{mun03a} and their counterparts from the 2MASS catalog. 
The rms dispersion in the offsets was 0\farcs25. We conclude that the 
positions of individual X-ray sources are accurate to 0\farcs3 with 
90\% confidence. 

The image of the 20\arcsec$\times$20\arcsec\ around \sgrastar\ is
displayed in Figure~\ref{fig:img}. The location of \newsource\ is 
$\alpha$=266\fdg41680, $\delta$=--29\fdg00861 ($\pm$0\farcs3; J2000).
Inspection of the figure reveals that the appearance of the transient 
was accompanied by a factor of two increase in the flux of the diffuse 
X-ray emission 2\arcsec\ south of the transient. The region is indicated by 
the white ellipse.

\begin{figure}
\centerline{\epsfig{file=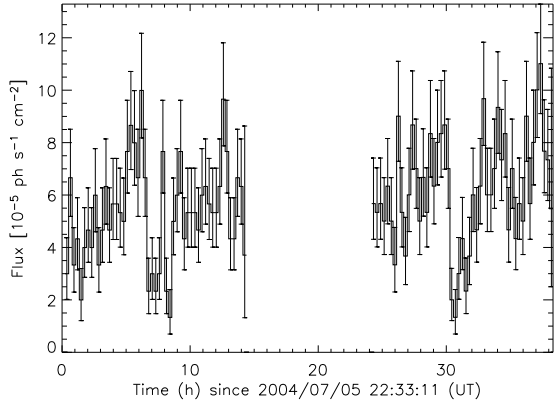,width=0.95\linewidth}}
\caption{
Flux, in units of \phcms, as a function of time from \newsource\ on 
2004 July 5--7. The source is clearly variable, with three minima
in the flux at about 0, 8, and 31 h. A Fourier analysis indicates
that these minima occur with a period of 7.9~h. 
}
\label{fig:lc}
\end{figure}

\begin{figure}
\centerline{\epsfig{file=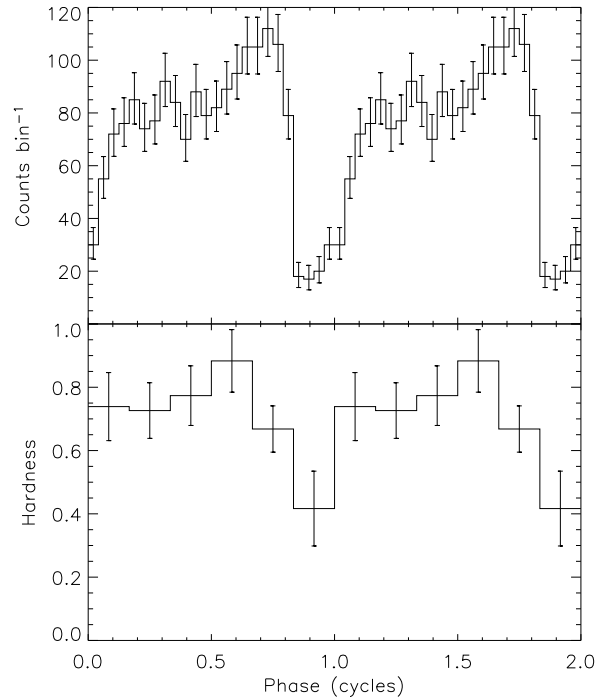,width=0.95\linewidth}}
\caption{
The folded profile of the 7.9 h modulation in the X-ray light curve
(top panel), along with the variation in the hardness ratio throughout
the cycle (bottom panel). We have repeated a single cycle twice in both 
panels. The hardness is defined as the ratio of counts
in the 2--6 keV and 6--8 keV energy bands. The softening of the spectrum
at the time of the dip in the lightcurve is only signifcant at the 93\% 
confidence level, and should be confirmed by more sensitive observations. 
}
\label{fig:prof}
\end{figure}

\subsection{Transient Properties}

In order to understand the nature of \newsource, we analyzed 
the light curve and spectrum for \newsource\ in the 
0.5--8.0~keV band using the acis\_extract routine from the Tools for 
X-ray Analysis (TARA)\footnote{www.astro.psu.edu/xray/docs/TARA/}
and CIAO version 3.0.2.
From each observation, we first extracted events associated with the source 
from a circular region that enclosed 90\% of the 
point spread function. The region had a radius of $\approx$1\arcsec. 
Then we extracted background event lists for each observation 
from larger circular regions centered on \newsource, excluding from the 
event list the point sources and discrete filamentary features in the 
field. We chose the size of the background region to include 
$\approx 1400$ total counts from the full set of observations.

In Figure~\ref{fig:lc}, we display the flux as a function of time during 
the observations on 2004 July 5--7. Three minima are evident in the 
light curve at about 0, 8, and 31~h. As reported in \citet{mun05},
we searched these observations for periodic variability using the Rayleigh 
statistic \citep[$Z_1^2$;][]{buc83}. The the power spectrum contains a 
strong signal with a period of 7.9~h  and $Z_1^2 = 175$, and its 
first harmonic with $Z_1^2 = 120$. The exact significance of this signal
is uncertain, because there may be red noise in the power spectrum. 
However, \xmm\ observations in 2004 August detected similar dips
with the same 7.9~h period \citep[][D. Porquet \etal, in prep.]{bel05}
making us confident that this signal represents a stable periodicity in
the source, such as its orbital period. 

In order to further explore the nature of the modulation, in the top panel 
of Figure~\ref{fig:prof} we display the 2--8 keV count rate folded about the 
7.9 h period. The most obvious feature in the folded profile is the 
dip mentioned above, during which the flux drops by 75\%. 
We also computed the ratio of counts in the 2--6 and 6--8 keV energy bands,
which is displayed in the bottom panel of Figure~\ref{fig:prof}.
This dip appears to be accompanied by a 
slight softening of the spectrum, although a $\chi^2$ test only allows
us to reject the hypothesis that the hardness is constant at the 93\%
confidence level. As we discuss in \S4, the dips in the light curve
of \newsource\ probably result from structures in the 
outer accretion disk that obscure the central X-ray emitting region.

\begin{figure}
\centerline{\epsfig{file=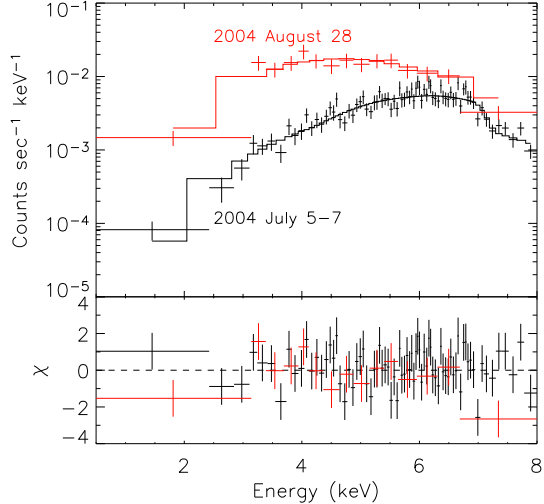,width=0.85\linewidth}}
\caption{
Spectra of \newsource\ obtained on 2004 July 5--7 and August 28.
{\it Top Panel:} The 
crosses denote the observed detector counts as a function 
of energy, so that the shape of the instrinsic spectrum is convolved with the 
instrument response. The solid lines indicate the detector counts 
predicted by the best-fit model. The
flux between 7--8 keV is very similar in both observations, however
the slope at lower energies is dramatically different. The smaller 
flux at low energy in the July observations probably results from 
excess absorbing material that is located within the binary system.
{\it Bottom Panel:} The difference between the counts observed and 
those predicted by the best-fit model, in units of residuals divided 
by the uncertainties.}
\label{fig:spectra}
\end{figure}

\begin{deluxetable}{lcc}
\tablecolumns{3}
\tablewidth{0pc}
\tablecaption{Spectral Properties of \newsource\label{tab:spectra}}
\tablehead{
\colhead{} & \multicolumn{2}{c}{Value}\\
\colhead{Parameter} & \colhead{2004 Jul} & \colhead{2004 Aug}
} 
\startdata
$N_{\rm H}$ ($10^{22}$ cm$^{-2}$) & $6^{+2}_{-1}$ & 6\tablenotemark{a} \\
$f_{\rm pc}$ & $0.92^{+0.01}_{-0.02}$  & $>0.5$ \\
$N_{\rm H,pc}$ ($10^{22}$ cm$^{-2}$) & $59^{+9}_{-8}$	& $5^{+4}_{-2}$ \\
$\Gamma$ & $0.0^{+0.6}_{-0.2}$ & $1.2^{+0.8}_{-0.4}$ \\
$N_\Gamma$ ($\times10^{-4}$ ph cm$^{-1}$ s$^{-1}$ keV$^{-1}$) &
				$1.0^{+2.5}_{-0.5}$ &
				$6^{+20}_{-3}$ \\
$\chi^2/\nu$ & 76/72 & 8/11 \\
$F_{\rm X}$ ($10^{-12}$ \ergcms) & $0.9$ & $2.2$\\
$uF_{\rm X}$\tablenotemark{b} ($10^{-12}$ \ergcms) & $1.2$ & $3.3$\\
$L_{\rm X}$\tablenotemark{c} ($10^{34}$ \ergs) & $3.7$ & $3.4$
\enddata
\tablenotetext{a}{The column density for the 2004 August observation was 
set to the value toward \sgrastar.}
\tablenotetext{b}{We define $uF_{\rm X}$ as the de-absorbed flux after 
correcting for interstellar absorption.}
\tablenotetext{c}{We define $L_{\rm X}$ as the luminosity after correcting 
for both interstellar and local absorption.}
\tablecomments{Uncertainties are 1$\sigma$ for a single parameter of 
interest ($\Delta\chi^2 = 1$). We report the 2--8~keV fluxes and 
luminosities, because this is the band in which most of the flux is 
observed. The bolometric values are probably less than a factor of 
2 higher.}
\end{deluxetable}

\begin{figure}
\centerline{\epsfig{file=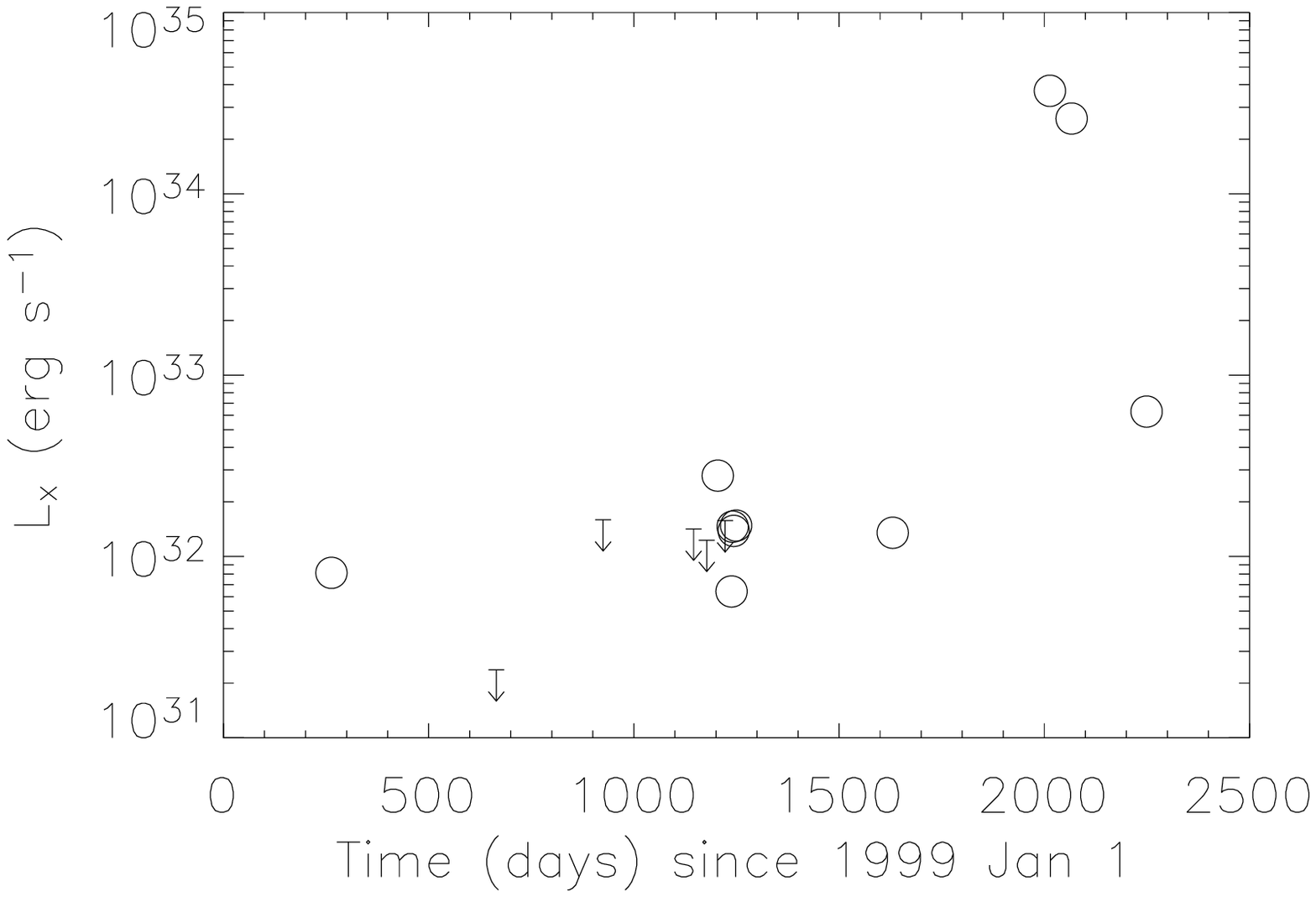,width=0.95\linewidth}}
\caption{
The luminosity of \newsource\ as determined from observations with
\chandra. Prior to 2004, the source was no brighter than 
$3\times 10^{32}$~\ergs. Observations in 2004 July and August reveal that 
the source has varied by a factor of $\approx 2$
about a mean luminosity of $5\times10^{34}$~\ergs. \xmm\ observations
in 2004 March and August reveal the source at a similar luminosity
\citep[][D. Porquet \etal, in prep]{bel05}.
}
\label{fig:hist}
\end{figure}

\begin{deluxetable*}{lccccc}[thb]
\tabletypesize{\scriptsize}
\tablecolumns{6}
\tablewidth{0pc}
\tablecaption{Spectral Properties of the Diffuse Emission\label{tab:diff}}
\tablehead{
\colhead{Date} & \colhead{$N_{\rm H}$} & \colhead{$kT$} & 
\colhead{$K_{\rm EM}$} & \colhead{$F_{rm X}$ [2--8~keV]} & 
\colhead{$L_{\rm X}$ [2--8 keV]} \\
\colhead{} & \colhead{($10^{22}$ cm$^{-2}$)} & \colhead{(keV)} &
\colhead{(cm$^{-6}$ pc)} & \colhead{(\ergcms)} &
\colhead{(\ergs)} 
} 
\startdata
\cutinhead{Varying $kT$}
1999 Sep to 2003 Jun & $8^{+2}_{-2}$ & $1.5^{+0.6}_{-0.4}$ & $9^{+13}_{-4}\times10^{-3}$ & 
  $0.5\times10^{-13}$ & $1.1\times10^{33}$ \\
2004 Jul to Aug & 8\tablenotemark{a} & $3^{+3}_{-1}$ & $9\times10^{-3}$\tablenotemark{a} & 
  $1.0\times10^{-13}$ & $2.3\times10^{33}$ \\
\cutinhead{Varying $K_{\rm EM}$}
1999 Sep to 2003 Jun & $8^{+2}_{-2}$ & $1.7^{+0.8}_{-0.4}$ & $8^{+10}_{-5}\times10^{-3}$  
& $0.5\times10^{-13}$ & $1.1\times10^{33}$ \\
2004 Jul to Aug & 8\tablenotemark{a} & 1.7\tablenotemark{a} & 
$16^{+21}_{-10}\times10^{-3}$ & $1.0\times10^{-13}$ & $2.3\times10^{33}$
\enddata
\tablenotetext{a}{These parameters were tied together in the two spectral
fits. The values of reduced chi-squared for each joint fit were 
$\chi^2/nu$=$6/37$ when allowing $kT$ to vary, and $\chi^2/nu$=$6/38$ when
allowing $K_{\rm EM}$ to vary.}
\end{deluxetable*}

Next, we extracted and modeled the spectrum of \newsource. We 
produced source and background spectra
from the respective event lists by computing the histogram over pulse 
height (energy). We then computed the effective area function at the 
position of \newsource\ for each observation. This was corrected to 
account for the fraction of the PSF enclosed by the extraction region. 
Finally, we estimated the detector 
response for the source in each observation using position-dependent 
response files that accounted for the corrections we made to undo 
partially the charge-transfer inefficiency \citep{tow02a}.
The mean flux did not change between the two observations on 
2004 July 5--7, so we summed the source spectra, and computed the 
average effective area and response functions weighted by the number
of counts from the two observations. 

We obtained enough photons to model the spectra from the observations on
2004 July 5--7 (1740 total counts) and 2004 August 28 (306 total counts). 
We subtracted the same background spectrum from the 2004 July and August 
spectra. The background contributed $<$3\% to the total flux. 
We grouped the source spectra so that each energy bin had at least 20 
counts. The spectra are displayed in Figure~\ref{fig:spectra}. We modeled 
the spectra using XSPEC version 11.3.1 \citep{arn96}. We initially modeled 
the spectrum a power-law absorbed by interstellar gas and dust. However, 
we found that for the longer observation, this model did not re-produce 
the 0.5--2.0 keV part of the spectrum and left residuals near the 
photo-electric edge of Fe at 7~keV. Therefore, we added a second absorption 
component that only affected a fraction of the emitting 
region. The free parameters in this model were the column of interstellar 
gas ($N_{\rm H,ISM}$), 
the column of the partial-covering absorber ($N_{\rm H,pc}$) and the 
fraction of the emitting region covered by this absorber ($f_{\rm pc}$), 
the photon index ($\Gamma$) and normalization ($N_\Gamma$) of 
the power law. The optical depth of dust was set to 
$\tau = 0.485 \cdot N_{\rm H}/(10^{22} {\rm cm}^{-2})$, and the halo area
to 100 times that of the PSF \citep{bag03}. 
The best-fit spectral parameters for the two epochs are listed in 
Table~\ref{tab:spectra}. We found that the interstellar absorption in 
the 2004 July spectrum is consistent with the value toward 
\sgrastar, $6\times10^{22}$~cm$^{-2}$. There was not enough signal
in the 2004 August observation to constrain all of the parameters, 
so we fixed the interstellar absorption to this value.

The change in the spectrum in Figure~\ref{fig:spectra} is the result of
an order-of-magnitude decrease in the column 
depth of the partial-covering absorber between 2004 July and August.
In contrast, the values of the photon indices from the two observations 
are consistent within their 2$\sigma$ uncertainties, so the intrinsic 
spectrum appears to change very little. Likewise, after accounting for 
the interstellar and partial-covering absorption, the inferred 
luminosity of \newsource, changes by only 10\% between 2004 July and
August.

Finally, in order to determine the quiescent luminosity of \newsource, 
we extracted the counts from the \chandra\ observations
during 1999--2003, and during February 2005. During 1999--2003, 
the region contained 485 total counts, of which the expected background 
contribution was 
371 counts. Although this excess flux could represent the quiescent emission 
from \newsource, two young, emission-line stars also lie 
within the 1\arcsec\ extraction region (IRS 33N and IRS 33E), 
and these may be X-ray sources. If we assume a 
a $\Gamma = 1.5$ power law spectrum (typical for a quiescent LMXB; 
e.g., Kong \etal\ 2002) absorbed by $6\times10^{22}$ cm$^{-2}$ of gas and
dust, we can place a rough upper limit
to the quiescent luminosity of \newsource\ using the observed 
net count rate of $1.8\times10^{-4}$ count s$^{-1}$ during 1999--2003. 
We find that the quiescent luminosity is 
$L_{\rm X} \la 7\times10^{31}$~\ergs\ (2--8 keV).

During the 2005 February observation,
we received only 13 counts from the location of \newsource, 5 of which 
should be background.  If the total count rate were the same as that 
before the outburst, $8\times10^{-4}$ count s$^{-1}$, the chance 
probability of receiving at least 13 counts in 5 ks from a Poisson 
distribution is only $5\times10^{-5}$.
Therefore, X-rays continuted to be produced by the outburst through 
2005 February. The corresponding luminosity is 
$L_{\rm X} \approx 6\times10^{32}$~\ergs.
The history of the luminosity of \newsource\ is displayed in 
Figure~\ref{fig:hist}.

\subsection{Diffuse Feature}

In order to study the properties of the diffuse X-ray emission that 
brightened coincident with the outburst of \newsource, we have 
defined an ellipse that encloses the excess counts seen during 2004 July.
This ellipse is displayed in Figure~\ref{fig:img}. It is centered at 
$\alpha$=266\fdg41629, $\delta$=-29\fdg00936 (1\farcs6 east and 
2\farcs7 south of \newsource), and its semi-major and semi-minor axes 
are respectively 3\farcs0 and 1\farcs6. The ellipse is somewhat 
larger than the extent of the most obvious brightening of the diffuse 
flux because we wanted to include photons in the wings of the \chandra\ 
PSF, which extends a factor of $\approx$4 beyond its 0\farcs5 core.

We judged the significances of the changes in flux by assuming 
that the numbers of counts follow a Poisson distribution. We
found that the count rate increased between 1999-2003 and 2004 July, from
$(2.08 \pm 0.05) \times10^{-3}$ count s$^{-1}$ to 
$(4.1 \pm 0.2)\times10^{-3}$ count s$^{-1}$. The increase in flux is 
significant at the $\approx$10$\sigma$ level. The count rate in 
2004 August, $(5 \pm 1) \times10^{-3}$ count s$^{-1}$, was consistent with 
that in 2004 July, and 3$\sigma$ larger than during 1999--2003. 
Finally, in 2005 February, the flux appeared to return to its value in 
1999--2003, with a count rate of $(2.1 \pm 0.7) \times 10^{-3}$ count s$^{-1}$.

\begin{figure}
\centerline{\epsfig{file=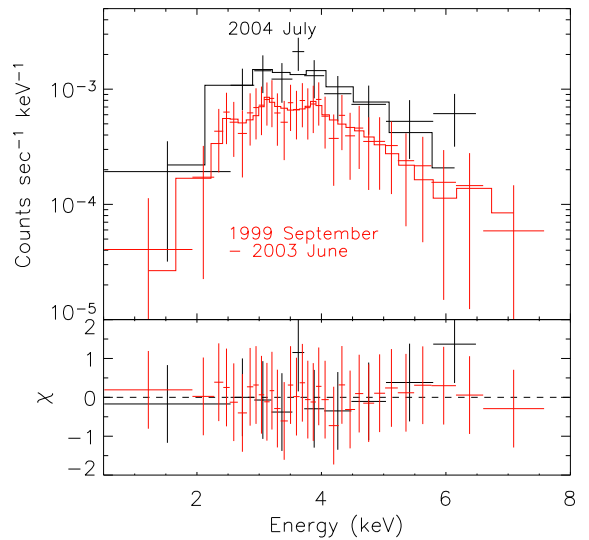,width=0.9\linewidth}}
\caption{
Spectra of the diffuse emission south-west of \newsource\ 
obtained on between 1999 September through 2003 June (red) and 
2004 July 5--7 (black). The intensity of the diffuse emission clearly
increases. However, there is not enough signal in the spectrum to 
determine what physical changes in the emitting region produced the 
brightening.}
\label{fig:diff}
\end{figure}


Next, we attempted to determine what physical changes occurred when the 
region brightened. We extracted the spectra of the diffuse emission within the
elliptical region from each observation prior to 2005, obtained 
the appropriate instrument response files \citep{tow02a}, and computed 
an effective area functions weighted by the number of counts in each 
pixel. We then computed the 
summed spectra and count-weighted average responses for two intervals, 
1999 September through 2003 June, and 2004 July through August.
For the background subtraction, we used the spectrum of the diffuse emission 
extracted from the ``close'' region 4\arcmin\ south 
of \sgrastar\ in \citet{mun04a}. The background contributed $<3$\% to 
the total spectrum. We grouped the spectra to have a minimum of 40 counts per 
bin between 0.5 and 8 keV. The spectra are displayed in 
Figure~\ref{fig:diff}. 

We modeled the spectra as a thermal plasma 
\citep{mew86} absorbed by interstellar gas and dust. 
The free parameters of the model were the absorption 
$N_{\rm H}$, temperature $kT$, and normalization (proportional 
to the emission measure $K_{\rm EM} = \int n_e n_H dV$). 
We found that the differences 
in the two spectra were consistent with changes in either $kT$ or 
$K_{\rm EM}$. We we list the best-fit paramters for both hypotheses 
in Table~\ref{tab:diff}.
If we assume that both $kT$ and $K_{\rm EM}$ varies, an $F$-test suggests 
that the change in $\chi^2$ has at least a 10\% chance of representing a 
random variation.
Unfortunately, beyond identifying a brightening in the 
diffuse emission, there is not enough signal to determine unambiguously
what physical change occurred in the region to produce the enhanced emission.
In \S4.2, we interpret the diffuse emission as scattered light from the
outburst of \newsource.


\section{Infrared Observations}

We searched for an infrared counterpart to 
\newsource\ using observations taken at the Keck Observatory. 
For each observation, we calculated the astrometry using the positions of 
known maser sources in the field, so the coordinates are accurate to 
10 mas \citep{ghe05}.

\begin{figure}
\centerline{\epsfig{file=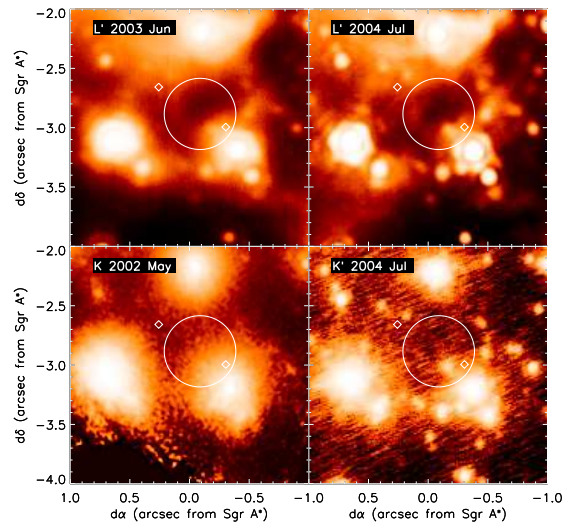,width=0.95\linewidth}}
\caption{$L^\prime$ ({\it top}) and $K/K^\prime$ ({\it bottom}) images 
of the 2\arcsec\ by 2\arcsec\
field around \newsource\ taken in 2003 ({\it left}) and 2004 
({\it right}) with Keck (see text for details). The position of 
the transient X-ray source is indicated with the circle of 0\farcs3
radius, and the mean positions of the two radio features are indicated 
with diamonds. Several stars are present within the error 
circle of the transient, although only the $K^\prime$ image from 2004 July
is sensitive enough to see all of them. Therefore, we cannot identify 
the true infrared counterpart of \newsource. The faintest stars in the images
have $L^\prime \sim 15$ and $K^\prime \sim 17$, which we take as the 
upper limits to the magnitude of the counterpart.}
\label{fig:irzoom}
\end{figure}

To search for the transient in outburst, we used images at 
L$^\prime$ and $K^\prime$ that were 
taken in 2004 July using the NIRC-2 camera behind the newly-commissioned
laser guide star adaptive optics system. 
To establish whether there were variable infrared sources within the 
error circle of \newsource, we compared these images from to a $K$ speckle 
image taken in 2002 May with NIRC 1 and an $L^\prime$ image taken in 
2003 July with NIRC-2 behind natural guide star adaptive optics. 
Unfortunately, these earlier images did not benefit
from the increase in quality afforded by the laser guide star system, and are
not as sensitive as the images taken in 2004 July.
The 2\arcsec\ by 2\arcsec\ field around the transient from each image
is illustrated in Figure~\ref{fig:irzoom}. In both images, the position 
of \newsource\ is indicated with a 0\farcs3 circle (90\% confidence), 
and the average locations 
of the two lobes of the radio jet are illustrated with diamonds \citep{bow05}.

\begin{figure}
\centerline{\epsfig{file=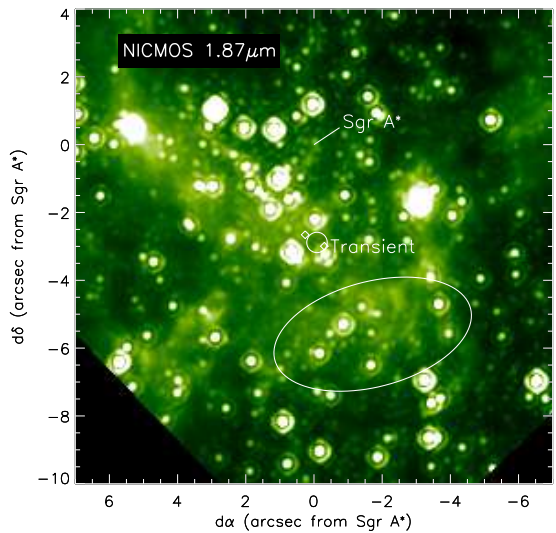,width=0.95\linewidth}}
\caption{
An HST NICMOS image taken in 1998 with the narrow 1.87 $\mu$m filter.
The filter is sensitive to Paschen-$\alpha$ emission from Hydrogen,
which is evident in this image as bands of diffuse emission. 
We have marked the relative locations 
of \sgrastar, the transient, the two radio features (diamonds), and the 
region of enhanced diffuse X-ray emission (large ellipse). The region of 
enhanced diffuse X-ray emission is coincident with a dense region of ionized
gas that is evident by its Hydrogen emission lines.
}
\label{fig:paalpha}
\end{figure}

\begin{figure}
\centerline{\epsfig{file=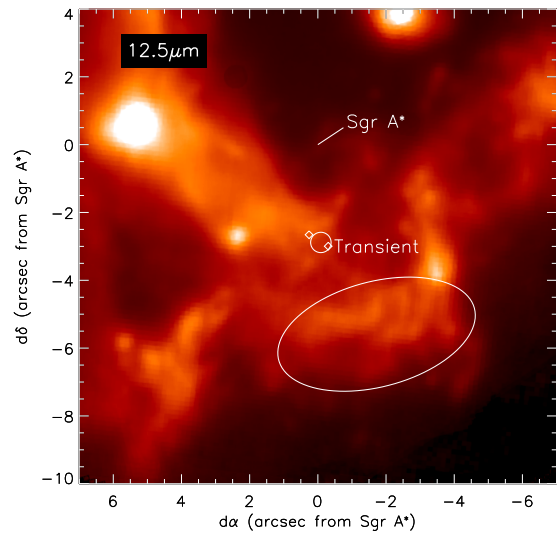,width=0.95\linewidth}}
\caption{
A mid-infrared (12.5 $\mu$m) image of the Galactic center taken with
the MIRLIN camera on Keck I in 1998. The image is dominated by thermal
emission from dust. We have marked the relative locations 
of \sgrastar, the transient, the two radio features (diamonds), and the 
region of enhanced diffuse X-ray emission (large ellipse). Two things 
are notable. First, dust emission to the east of the transient is 
particularly strong, coincident with the brighter of the two radio features.
This suggests that the radio feature is produced where a jet interacts
with the ISM. Second, as in the Paschen-$\alpha$ image, the region of 
enhanced diffuse X-ray emission lies on a bright section of dust emission.
}
\label{fig:midir}
\end{figure}

There are clearly several faint infrared sources within the 90\%
error circle of \newsource\ in the images taken in July 2004 (right panels, 
Fig.~\ref{fig:irzoom}). The brightest of these, at the southwest edge, 
has $K^\prime = 15.3$ and $L^\prime = 13.3$. Although the $K$ image
from 2003 is not sensitive enough to reveal this source, it does appear
to be present at the same intensity in $L^\prime$ in 2003. 
There may also be several fainter sources near the detection threshold of 
$K^\prime \sim 17$ in the $K^\prime$ image from July 2004. Again,
the $K$ images from 2003 are not sensitive enough to determine whether 
any of these brightened. Therefore, we are not able
to unambiguously identify the infrared counterpart to \newsource,
but can can rule out with 90\% confidence that its infrared counterpart 
has $K<15$.

Finally, we note that the $K^\prime$ and $L^\prime$ images extend 
no more than 4\arcsec\ south of \sgrastar, and do not cover the region 
where the diffuse X-ray emission brightened. Therefore, we have obtained
images with wider fields-of-view in order to search for gas and dust 
that might have contributed to the brightening of the diffuse emission.
In Figure~\ref{fig:paalpha}, we present a 1.87 $\mu$m image of the 
Galactic center taken in 1998 August with the Near Infrared Camera and 
Multi-Object Spectrometer (NICMOS) 
aboard the {\it Hubble Space Telescope} \citep[][]{sco03}. 
The astrometry was registered to 0\farcs1 by matching the brightest 
stars to sources in the 2MASS catalog.
The filter used is sensitive to Paschen-$\alpha$ emission from
Hydrogen, which can be seen as bands of diffuse emission in the image.
We also present a 12.5 $\mu$m image taken with the MIRLIN 
camera on Keck II during 1998 June. We computed the astrometry using the 
positions of IRS 3, IRS 7
and IRS 22 in \citet{gen00}, and estimate that the positional uncertainty
is also about 0\farcs1. This image is dominated by thermal emission from dust.

In both images, the ellipse denotes the region that brightened in X-rays.
It is clearly coincident with bright emission from ionized gas and dust.
The diamonds denote the two lobes of the jet \citep{bow05}. 
The brighter, eastern lobe 
is coincident with prominent dust emission in Figure~\ref{fig:midir}.
The same appears to be true for the ionized gas in Figure~\ref{fig:paalpha},
although the prominence of the stars in the 1.87 $\mu$m image 
makes this much less obvious.

\section{Discussion}

As stated in \citet{mun05}, our X-ray and infrared observations of 
\newsource\ suggest that it is a LMXB that is viewed nearly edge-on. 
Now that we have described the X-ray and infrared observations in 
greater detail, it is worth discussing how this conclusion was
reached.

First, the partial-covering absorption in the spectrum and
the 7.9~h modulation in the light curve suggests that
we observe \newsource\ nearly edge-on \citep[e.g.][]{par86}. The geometric 
arrangement that leads to the partial-covering absorption is as follows. 
Most of the X-ray emission originates in a region a few tens of kilometers 
across, consisting of the inner accretion 
disk and, for neutron star systems, a boundary layer where the accretion
flow is halted. The total radius of the accretion
disk is $\ga 10^5$ km, and, at a minimum (i.e., 
ignoring radiation-induced warping, etc.), the thickness of the 
accretion disk should increase as $H \approx \alpha R$, where 
$\alpha \sim (10^{-3} - 0.1)$ \citep[e.g.,][]{fkr95}. Therefore, the outer 
accretion disk has $H \ga 10^2$ km, and is easily thick enough to obscure the 
main X-ray emitting region for a system with an inclination $>$75\degree.
A fraction of the X-ray emission also originates in a 
corona of hot plasma above the accretion disk. The scale height of this 
emission is 
large enough that part of it is visible even from systems observed 
nearly edge-on, as is thought to be the case for several LMXBs 
that are referred to as accretion disk corona sources \citep{par00, kal03}.
The partial-covering absorption 
(Fig.~\ref{fig:spectra} and Tab.~\ref{tab:spectra}). 
could be produced because the upper layers 
of the outer accretion disk are not quite optically thick to electron 
scattering ($\tau \sim 1$).

Dips, similar to those seen in Figure~\ref{fig:lc}, are caused by discrete 
structures that rise above the outer accretion disk 
and obscure the X-ray emission for a small fraction of the binary orbit. 
Such structures form, for example, at the point where the material lost
by the companion star first impacts the accretion disk.
Since most of the absorption results from the incident X-rays photoionizing 
the intervening material, the dips should be most prominent at low-energies.
This is the case in about half of all edge-on LMXBs \citep{bc99,hom03,uno97}. 
However, in the other half the depths of the dips are indepedent of energy
\citep{whi84,par99,iar01}, as in Figure~\ref{fig:prof}. 
This is probably because either (1) 
the dips result from energy-independent electron scattering by 
highly-ionized or metal-deficient material, or 
(2) the mean energy of the X-rays emitted from the corona decreases as 
a function of height, so that much of the cooler flux remains unobscured 
during the dips. Should the evidence that we find for softening during the 
dips in Figure~\ref{fig:prof} be confirmed, some combination of these 
options can be constructed to reproduce the spectral evolution of the dip.

The fact that \newsource\ is a compact object accreting from a 
low-mass companion is indicated by the 7.9 h orbital 
period of the binary (Fig.~\ref{fig:lc}), and by the faintness of the 
infrared companion (Fig.~\ref{fig:irzoom}).
The short orbital period can accommodate a mass donor with a radius 
of $\approx 0.8$~$R_\odot$ (Frank, King, \& Raine 1995, eq. 4.10). 
The only high-mass stars that are this compact are in the Wolf-Rayet phase. 
For $D=8$ kpc and $A_K = 3.2$ (Reid \etal\ 1999; Tan \& Draine 2003) 
a Wolf-Rayet star would have $K = (9-14)$, and would have been 
easily detectable in our Keck images. In contrast, if the few LMXBs 
that have been monitored in the infrared during their outbursts were 
placed at the Galactic center, they would have 
had peak intensities of $K \approx (15-17)$
\citep{jai01,cha03,bb04}. The fainter LMXBs would have been barely detectable
in our 2003 Keck images. Therefore, the short orbital period and the 
lack of a counterpart with $K < 15$ in Figure~\ref{fig:irzoom} indicates 
that the mass donor in \newsource\ is a low-mass star that over-fills its
Roche lobe.

The nature of the compact object is not yet clear. We did not
observe either thermonuclear bursts from the surface of the compact object, 
or coherent pulsations that can be naturally associated with a 
spin period. Either of these would indicate that the source is a neutron 
star. However, the lack of theses signals is not surprising,
because the recurrence times of bursts 
are often longer than 100~ks and the time resolution of the ACIS data was 
too coarse to 
detect pulsations faster than 10~s. On the other hand,
bright radio emission is much more common from black hole LMXBs than 
neutron star ones \citep{fk01}, which suggests that the
primary in \newsource\ could be a black hole.

\subsection{A Faint X-ray Transient}

One unusual aspect of the outburst 
from \newsource\ is that it is quite faint. The history of its X-ray 
luminosity between 1999 and 2004 is displayed in 
Figure~\ref{fig:hist}. During 1999--2003, \chandra\ detected a 
marginally-significant excess in the source counts within 1\arcsec\ of 
\newsource, with an average luminosity of 
$\sim 10^{32}$~\ergs. This emission is probably from nearby young, 
emission line stars. Therefore, we consider $10^{32}$~\ergs\ as the upper
limit to the luminosity of \newsource\ during that time period. This is 
typical for an LMXB in quiescence. In contrast, the highest luminosity 
observed from \newsource\ with \chandra\ was only $4 \times 10^{34}$~\ergs. 
This luminosity is well below those at which 
transient LMXBs are typically detected in outburst \citep[e.g.][]{cam98}.
The low luminosity is surprising in the context of the disk-instability 
models that
are typically used to explain the outbursts of LMXBs, which predict that
the entire accretion disk is be disrupted, leading to an outburst with 
$L_{\rm X} > 10^{37}$~\ergs\ \citep[e.g.,][]{kr98}.
However, three observational selection effects could contribute to the low
peak luminosity of \newsource.

First, transient LMXBs have
traditionally been identified with wide-field monitoring instruments
that only have sensitivities of $\ga 10^{-10}$~\ergcms\ \citep{lev96,jag97},
or $10^{36}$~\ergs for the Galactic center distance. There are
few accreting black holes and neutron stars within 2~kpc of Earth, so there
is a strong selection effect against finding transients this faint.

Second, only a few sensitive X-ray observations of the Galactic center 
have been obtained within the last year, so we may have missed the
peak of the outburst. The
first observation of the source in outburst was taken by \xmm\ in 2004 
March \citep[][D. Porquet \etal, in prep]{bel05}.
They report that the source had a comparable luminosity to that at which 
we detected the source with \chandra\ in 2004 July and August. In
between time, the {\it Rossi X-ray Timing Explorer} carried out
scanning observations of the Galactic center with Proportional Counter 
Array \citep{mar02}.
These allow us to put an upper limit of $3 \times 10^{36}$~\ergs\ on 
the intensity of the source during 2005. This upper limit is 
at the low end of the luminosities of outbursts often seen from LMXBs.

Finally, as mentioned above, we observe \newsource\ along the plane of its 
binary orbit, so it is likely that the outer accretion disk obscures most
of the X-ray emitting region. 
These facts motivate us to search for an independent constraint on the
X-ray luminosity of the transient outburst. Fortunately, the apparent 
detection of scattered X-ray emission from the transient outburst in 
Figure~\ref{fig:img} provides us with just such a constraint.

\subsection{The Light Echo}

The enhancement in the diffuse X-ray emission is coincident with part of a
well-known ridge of dust and ionized gas, referred to as the ``Minispiral''
(Fig.~\ref{fig:midir} and \ref{fig:paalpha}).
Although the brightening in diffuse X-rays could represent either scattered 
X-rays or a mechanical outflow shocking against the surrounding 
interstellar medium, the morphology and energetics of the observed 
X-rays makes the first mechanism appear more likely. First, the 
diffuse region is separated from the central source by at least 4 light-months 
and the outburst of \newsource\ started after June 2003, so 
any material that impacted the Minispiral must have been traveling faster 
than $0.3c$.
Second, if we interpret the brightening of the diffuse X-rays
as an increase in the density of the emitting plasma, then the 
implied energy input is $\Delta U = (\Delta n) kT \sim 10^{42}$~erg
(Table~\ref{tab:diff}).
Therefore, the power required over six months is $10^{35}$~\ergs. 
Both of these conditions could be fulfilled by a radio 
jet \citep[see \S4.2 and][]{bow05}. However, the axis of 
the observed jet is oriented 
about 45\degree\ from the center of the brightening of the diffuse 
emission. Moreover, the diffuse emission has an extent of $\approx$3\arcsec, 
which is vastly more extended than the radio features. These
facts make it seem unlikely that the jet is responsible for 
the diffuse X-ray emission. Instead, we propose that the enhancement 
in the diffuse emission 
is produced by X-rays from \newsource\ that are scattered by electrons
in the Minispiral. 

The ionized gas from this ridge has been extensively studied in 
radio continuum at 13 mm \citep{zg98} and 6cm \citep{lc83}; hydrogen 
emission from the H92$\alpha$ \citep[3.6 cm][]{rg93}, Pa$\alpha$ 
\citep[1.87 $\mu$m][]{sco03}, and Br$\gamma$
\citep[2.16 $\mu$m][]{pau04} electronic transitions; 
and [Ne II] emission \citep[12.8 $\mu$m][]{las91,vd00}.
The ridge is obviously more extended than the brightening of the diffuse 
emission (Fig.~\ref{fig:paalpha}), which raises the question of why only 
a small fraction of it has been illuminated. Careful studies of the velocity 
of the gas in the 
Minispiral have indicated that it is composed of several 
kinematic features \citep[][Paumard, Maillard, \& Morris 2004]{vd00}. 
Although the complexity of the region precludes any conclusive associations,
the enhancement in X-ray flux is coincident with a section of the 
``Northern Arm'' that has a high velocity toward 
us \citep[200 km s$^{-}$; see][]{pau04}. We 
suggest that this is the only region that has 
brightened because it is closest to \newsource. 

The scattered flux ($F_{\rm scat}$) associated with the outburst 
of \newsource\ depends on the luminosity ($L_{\rm X}$) and distance 
($D$) of the source, the solid angle ($\Omega$) and optical depth 
($\tau$) of the scattering region, and the angle ($\theta$) through which
photons are scattered:
\begin{equation}
F_{\rm scat} = L_{\rm X} \frac{f(\theta)}{4\pi D^2} (1 - e^{-\tau}) \frac{\Omega}{4\pi}.
\label{eq:scat}
\end{equation}
The function $f(\theta)$ depends on the scattering process; for Thompson 
scattering $f(\theta) = 0.75(1+\cos^2\theta)$. 
The solid angle $\Omega$ depends 
on the distance between the source 
and the scattering region, and the size of the scatterer. 
The projected separation between the two is
$\approx$2\arcsec, so if $\theta$ is the angle between our line of sight
and the line connecting the source and the scatterer, the true distance is
$d \approx 0.1 (\sin\theta)^{-1}$~pc. The brightening of the diffuse flux is 
contained in a roughly elliptical region no larger than 
3\farcs0$\times$1\farcs6, so we estimate that the projected area of the 
scattering region with respect to the source is $A \la 0.02$~pc$^{2}$.
Therefore, the solid angle of the scatterer is 
$\Omega/4\pi = A/(4\pi d^2) \approx 0.2 \sin^2\theta$. 

The scattering is most likely caused by electrons.\footnote{Dust only
scatters photons by a few arcminutes, so the hypothesis that the 
diffuse X-rays are scattered by dust and would require \newsource\
lie $\sim$100 pc from the Minispiral, which in turn would imply
that its intrinsic luminosity is $L_{\rm} \sim 10^{42}$\ergs. We also see no 
evidence for fluorescent emission from metals, but this could result from 
the poor signal-to-noise in the spectrum of the diffuse emission.} 
The electron 
density in the region of enhanced diffuse X-ray emission 
has not been measured directly, but the average electron 
density has been determined over the entire Northern Arm by \citet{sco03} 
by comparing the Pa$\alpha$ and H92$\alpha$ fluxes. 
We use their value of $n_{\rm e} \sim 10^{4}$ cm$^{-3}$ to estimate the 
optical depth to electron scattering.
Assuming that the depth of the scattering region is similar to its major 
axis length, $l \approx 0.2$~pc, the column density is 
$N_{\rm e} = n_{\rm e} l \approx 6\times10^{21}$~cm$^{-2}$. The 
optical depth to Thompson scattering is then
$\tau_{\rm T} = \sigma_{\rm T} N_{\rm e} \approx 0.004$.

We have measured the scattered flux, 
$L_{\rm scat} = 1.6\times10^{-13}$~\ergcms\ (2--8 keV; Table~\ref{tab:diff}), 
so we can solve for the intrinsic luminosity of \newsource\ using
Equation~\ref{eq:scat}. If we assume $\theta = 90$\degree, we find 
$L_{\rm X} \approx 2\times10^{36}$~\ergs.
We consider this to be a conservative estimate of the 
intrinsic luminosity, because choosing a smaller value for $\theta$ 
or assuming a smaller depth and projected area for the scattering region
would result in an inferred luminosity that is a factor of several higher, 
up to $L_{\rm X} \sim 10^{37}$~\ergs. 

Therefore, the peak luminosity of \newsource\ must have been at least
$\sim 100$ times larger than the values observed with \chandra\ 
(Table~\ref{tab:spectra} and Figure~\ref{fig:hist}). As mentioned above, 
{\it RXTE} PCA observations would have detected an outburst larger than 
$3\times10^{36}$~\ergs, and therefore conceivably could have missed
\newsource\ at its peak luminosity. However, the timing and 
morphology of the diffuse emission argue that the peak of the outburst
should have been detected with \xmm. The region of diffuse X-ray emission 
remained bright for at least two months between 2004 July 5 and August 28 
(\S2.2), so 
the luminous portion of the outburst must have lasted at least this long. 
The peak of the outburst had 
to have occurred $d/c \approx 4\sin^{-1}\theta$ months prior to 
the \chandra\ observations. This places the peak of the outburst during 
2004 March and April, when \xmm\ observed the source. At that time the 
flux was similar to that in our \chandra\ observations 
\citep[][D. Porquet \etal, in prep.]{bel05}.

Therefore, it seems likely that the flux measured from the location of 
\newsource\ by \chandra\ and \xmm\ 
is only a small fraction of its total output. Indeed, 
observations of edge-on LMXBs often suggest that their 
intrinsic luminosities are significantly larger than would be inferred 
from their observed X-ray fluxes. 
For instance, based on the strength of oxygen emission lines in the \xmm\ 
Reflection Grating Spectrometer and \chandra\ High-Energy Transmission 
Grating spectra of 2S 0921-63, \citet{kal03} suggest that 
only $\sim 30$\% of its total X-ray flux is transmitted toward the observer.
Similarly, based on the low observed X-ray to optical flux of the
accretion disk corona source X~1822--731, 
\citet{par00} suggest that we observe only 5\% of its total flux. 
Our measurements of \newsource\ suggest that an even smaller fraction,
$\sim$1\%, of the total flux is observed. 

If indeed the vast majority
of the flux from \newsource\ is obscured by the accretion disk, then it
could help explain the fact noted by \citet{nm04} that no confirmed black hole
LMXB is known with an inclination larger than 
75\degree.\footnote{It has been proposed that 4U 1755--33 is a black hole
with $i$$>$75\degree\ \citep{whi84}, although no mass function has yet 
been measured to confirm the nature of the compact object.}
If only 1\% of the total flux can be detected
from a black hole LMXB observed edge-on, then almost all of the black 
hole transients in \citet{fk01} would have apparent 
$L_{\rm X} \la 10^{36}$~\ergs, and would have been almost undetectable by 
the all-sky monitors on \bepposax\ and \rxte\ \citep{lev96,jag97}. 
If \newsource\ contains a black hole primary, then it could be
the first such system observed edge-on. 

\subsection{Radio Jets from a Faint X-ray Transient}

VLA observations by \citet{bow05}
revealed two new radio sources that appeared coincident with the X-ray
outburst of \newsource. The X-ray source was located 
on the line between the two radio features, which suggests that they are 
produced by synchrotron emission from a jet launched by the X-ray source. 
The peak intensity was observed in 2004 March, with $S_{\nu} = 90$ mJy,
or $L_{\rm R} \approx 4\pi D^2 \nu S_{\nu} \approx 2\times10^{32}$ \ergs
at 43 GHz. In 2004 July, only the eastern feature was detected, with a 
flux density of $\approx 45$ mJy at 43 GHz, or 
$L_{\rm R} \approx 1\times10^{32}$ \ergs. The intensity varied by 
$\approx 20$\% from day-to-day. The spectrum of the emission is uncertain,
because even when measurements were made at multiple frequencies, VLBI 
observations revealed that the sources were resolved, and therefore each 
frequency samples a different spatial scale from the jet.
It is also noteworthy that, unlike the relativistically-expanding jets often 
seen from LMXBs \citep{fen04}, there was no proper motion of the radio 
sources along the jet axis 
(although there was some perpendicular to that axis; see Bower \etal\ 2005
for further discussion). Therefore, the radio features probably 
formed where the jet impacted the interstellar medium. Indeed, the 
mid-infrared image in Figure~\ref{fig:midir} reveals a significant
amount of dust that is near in projection to \newsource, particularly
at the location of the brighter, eastern radio feature.

The radio luminosity of \newsource\
is unusually large compared to the X-ray luminosity that we 
infer from the light echo, $L_{\rm X} \approx 2\times10^{36}$~\ergs. First, 
LMXBs typically are observed to produce extended radio jets only during 
outbursts with peak luminosities of $L_{\rm X} \ga 10^{37}$~\ergs\ 
\citep[e.g.,][]{fk01}. The only sources fainter than this that produced 
radio outbursts are SAX J1808.4-3658 ($L_{\rm X} = 5\times10^{36}$~\ergs) and 
XTE J1118$+$480 ($6\times10^{35}$~\ergs), and in neither 
case was a jet resolved. 
Second, the ratio of the peak X-ray to radio luminosities for LMXBs 
in \citet{fk01} is typically $L_{\rm X}/L_{\rm R} > 10^6$, whereas 
that from \newsource\ is $\la 10^{4}$. For comparison, 
the three LMXBs in \citet{fk01} with the brightest radio 
emission relative to their X-ray emission are XTE J1748--288 
($\log[L_{\rm R}/L_{\rm X}] = 5.8$), 
GRO J1655--40 ($\log[L_{\rm R}/L_{\rm X}] = 5.4$), 
and Cir X-1 in the 1970s [$\log(L_{\rm R}/L_{\rm X}] = 5.3$).
The unexpectedly bright radio emission from \newsource\ probably results from
the fact that 
the jet radiated with unusually high efficiently when it impacted
the surrounding ISM.

Finally, we can compute a lower limit to the power required to produce the 
radio-emitting jet by assuming that it contains only electrons and 
positrons, that 
these particles are in equipartition with the magnetic field, and that the
volume ($V$) of the emitting region is related to the time scale on which the 
jet is produced ($\Delta t$) by $V = 4\pi (c \Delta t)^3$ (Fender, Belloni, 
\& Gallo 2004)\nocite{fbg04}. 
If we assume 
that the slope of the radio spectrum is $\alpha = -0.75$, the minimum
energy is given by 
\begin{equation}
L_{\rm jet} = 2\times10^{35} (\Delta t_{\rm day})^{2/7} \nu_{\rm GHz}^{2/7} 
S_{\nu}^{4/7} D_{8~{\rm kpc}}^{8/7}~
{\rm erg~s}^{-1},
\end{equation}
where $\Delta  t_{\rm day}$ is the time over which the jet was launched
in days, $\nu_{\rm GHz}$ is the lowest frequency at which the radio 
emission was observed, $S_{\nu}$ is the flux density in mJy at that
frequency, and $D$ is the distance in units of 8~kpc 
\citep[][eq. 19.29]{lon94}. 
Using the peak radio flux of $S_\nu = 90$~mJy at $\nu_{\rm GHz} = 43$~GHz 
and assuming a time
scale of 1 day, we find that $L_{\rm jet} \approx 10^{37}$~\ergs. 
This power is comparable to the X-ray luminosity inferred from the 
diffuse X-ray light echo, which suggests that about half of the accretion 
energy is channeled into launching the radio jet. Several theoretical
arguments have suggested that LMXBs accreting at low rates 
release much of their energy as a jet 
\citep[e.g., Markoff, Falcke, \& Fender 2001,][]{mmf04,fbg04},
These observations of \newsource\ 
provide the most direct evidence yet that this is in fact true.

\section{Conclusions}

We have presented \chandra\ and Keck observations of a 
new transient X-ray source, \newsource, which is located only 0.1 pc from 
\sgrastar\ (Fig.~\ref{fig:img}). The presence of dips in the X-ray light 
curve that recur at the
7.9~h binary orbital period (Fig.~\ref{fig:lc}) and the lack of any infrared 
counterpart with $K^\prime < 15$ (Fig.~\ref{fig:irzoom})
indicates that this source is a low-mass X-ray binary. 
The peak flux from the transient is 
$F_{\rm X} = 4\times10^{-12}$~\ergcms\ (2-8~keV, de-absorbed), which
would imply a luminosity of only $L_{\rm X} = 3\times10^{34}$~\ergs 
(Fig.~\ref{fig:hist}). However,
the diffuse X-ray emission within 4 light-months of the source has also
brightened (Fig~\ref{fig:img}), probably because electrons in the 
ridge of gas and dust referred to as the the 
Minispiral (Fig.~\ref{fig:paalpha} and \ref{fig:midir}) have scattered 
light from the outburst. The intensity of the scattered flux suggests 
that the intrinsic luminosity is $L_{\rm X} \ga 2\times10^{36}$~\ergs. 

We have compared the energetics of the X-ray emission with those of transient
radio jets that were identified with the VLA \citep{bow05}. 
The brightness of the radio flux relative to the
X-ray flux suggests that the radio jets from \newsource\ radiate with 
unusually high efficiency, probably because electrons are accelerated
as the jet impacts the surrounding ISM. Moreover, we find that the X-ray 
luminosity is similar to the minimum energy required to power the radio 
jets, which provides the most direct evidence yet that LMXBs accreting
at low rates release most of their energy in the form of jets.

Future observations of this transient are important for several reasons.
First, such faint outbursts cannot be monitored with all-sky instruments
like that on \rxte. Therefore, the only way to learn about the duty 
cycles of such faint systems is to observe dense concentrations of LMXBs,
such as those at the Galactic center and in globular clusters, with 
\chandra\ and \xmm. Second, \chandra\ may be able to identify further 
brightening in the diffuse X-ray emission as the flux from the outburst
encounters more distant sections of the Minispiral. Finally, \chandra\ 
observations could reveal extended X-ray emission as the
radio jet impacts dense regions in the interstellar medium.

\acknowledgments
We thank G. B\'{e}langer for sharing information about the \xmm\ observations
of this transient before publication, G. Bower, F. Yusef-Zadeh, and
D. Roberts for sharing the results of the VLA observations, and C. Markwardt
for making the results of the \rxte\ Galactic bulge scans publicly available.
We also thank R. Wijnands for several insightful suggestions, and
H. Tananbaum for providing discretionary \chandra\ observations.
MPM was supported through a Hubble Fellowship grant (program number
HST-HF-01164.01-A) from the Space Telescope Science Institute, which is 
operated by the Association of Universities for Research in Astronomy,
Incorporated, Under NASA contract NAS5-26555.
WNB was supported by NSF CAREER Award 9983783. MRM, AMG, JRL, and SDH were
supported by NSF grant AST 99-88397 and the NSF
Science and Technology Center for Adaptive Optics, managed by the 
University of California at Santa Cruz under cooperative agreement AST 
98-76783, and the Packard Foundation. The W. M. Keck observatory is 
operated by the California Institute of Technology, the University of 
California, and NASA. The Observatory was made possible by the 
generous financial support of the W. M. Keck Foundation.

\end{document}